\documentclass[twoside,12pt]{article}
\usepackage{epsfig}

\def\Journal#1#2#3#4{{#1} {#2} (#4) #3 }

\newcommand{\be}{\begin{equation}}
\newcommand{\ee}{\end{equation}}
\newcommand{\bea}{\begin{eqnarray}}
\newcommand{\eea}{\end{eqnarray}}

\begin{document}

\title{ \vspace{1cm} A relativistic Glauber approach to polarization
transfer in $^4$He$(\vec{e},e'\vec{p})$ }
 \author{P.\ Lava,$^{1}$
J.\ Ryckebusch,$^1$ and B.\ Van Overmeire,$^1$ \\ $^1$Department of
Subatomic and Radiation Physics, Ghent University, Belgium} \maketitle
\begin{abstract}
Polarization-transfer components for $^4$He$(\vec{e},e'\vec{p})^3$H
are computed within the relativistic multiple-scattering Glauber
approximation (RMSGA). The RMSGA framework adopts relativistic
single-particle wave functions and electron-nucleon couplings. 
The predictions with free and various parametrizations for the medium-modified electromagnetic form factors are compared to the
world data.
\end{abstract}
In conventional nuclear physics nuclei are described in terms of
point-like protons and neutrons, interacting through the exchange of
mesons. It has been a long-standing and unresolved issue whether the
electromagnetic properties of bound nucleons differ from those of free
nucleons.  Exclusive $A(\vec{e},e'\vec{p})$ measurements have been put
forward as a tool to investigate the possible modifications attributed
to the presence of a medium.  In polarized electron free-proton
scattering,the ratio of the
electric ($G_E (Q^2=-q^{\mu}q_{\mu})$) to the magnetic ($G_M (Q^2)$)
Sachs form factors, can be extracted from \cite{arnold81}
\begin{equation}
\frac{P'_x}{P'_z} = -\frac{G_E (Q^2)}{G_M (Q^2) } \; \frac{2M_p}{E_e +
E_{e'}}\tan^{-1}\left(\frac{\theta_e}{2}\right) \; .
\end{equation}
Here, $q^{\mu}$ is the four-momentum transfer, $P'_{x}$ and $P'_{z}$
is the transferred polarization in the direction perpendicular to and
parallel with the three-momentum transfer, and $\theta_e$ the electron
scattering angle. For bound nucleons, deviations from the measured
ratio of $P'_x/P'_z$ from the above value ( thereby adopting
free-nucleon form factors) can indicate the existence of medium
modifications.  Finding signatures of medium modifications,
however, requires an excellent control over all those ingredients of
the $A(\vec{e},e'\vec{p})$ reaction process that are directly related
to the presence of a nuclear medium, such as final-state interactions
(FSI), meson-exchange currents (MEC) and isobar currents (IC). Of all
observables accessible in $A(\vec{e},e'\vec{p}) $, the transferred
polarization components have been recognized as the ones with the
weakest sensitivity to FSI, MEC and IC distortions. Recently,
$^{4}$He$(\vec{e},e'\vec{p})$ data have been reported
\cite{dieterich01,strauch03}, covering the range $0.4 \le Q^2 \le
2.6~$(GeV/c)$^2$. This kinematic regime may outreach the range of
applicability of optical-potential approaches for describing FSI
mechanisms. Indeed, given the highly inelastic and diffractive nature
of proton-nucleon scattering at proton lab momenta exceeding 1 GeV/c,
the use of optical potentials seems rather
unnatural and Glauber multiple-scattering theory provides a more
natural and economical description of FSI mechanisms
\cite{glauber70}. Recently, we developed an unfactorized and
relativistic version and dubbed it the relativistic
multiple-scattering Glauber approximation (RMSGA)
\cite{ryckebusch03}. In Ref.~\cite{lava04}, numerical calculations for
the polarization-transfer components in $^4$He$(\vec{e},e'\vec{p})$
are performed with both free and medium-modified electromagnetic form
factors. For the latter we used the predictions of the quark-meson
coupling model (QMC) \cite{lu98}.  In this contribution we use
alternative predictions of the chiral quark-soliton model (CQS)
\cite{smith04}. The CQS nucleon model puts more emphasis on the role
of the sea than the QMC framework. As a result,
the value of the magnetic moment remains practically unchanged.

In Figure ~1, the $^4$He polarization-transfer results are expressed
in terms of a double ratio $R$ \begin{equation} R =
\frac{(P'_x/P'_z)_{^4He}}{(P'_x/P'_z)_{^1H}} \; ,
\end{equation}
with the relativistic plane-wave impulse approximation (RPWIA) result
as baseline. Substituting the free forms factors with the CQS ones
reduces R. At $Q^2 \leq 1$ (GeV/c)$^2$ the reductions are
smaller than those observed for the computed values of the QMC model. At higher $Q^2$, both models
predict very similar effects. A better overall description of the data
is obtained with the medium-modified form factors.
\begin{figure}[tb]
\begin{center}
\begin{minipage}[t]{8 cm}
\epsfig{file=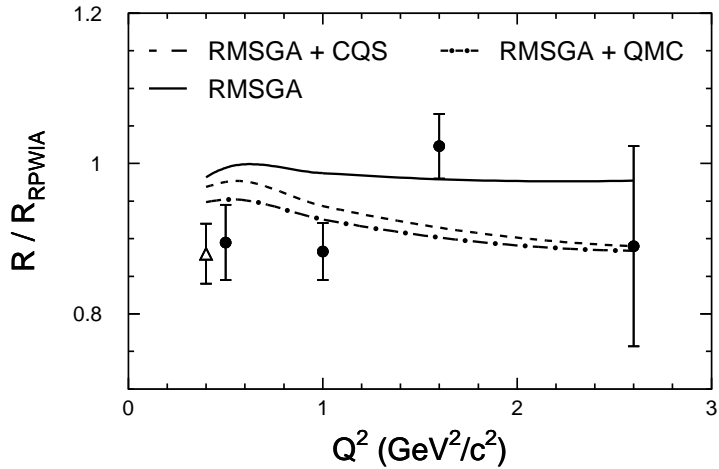,scale=0.5}
\end{minipage}
\begin{minipage}[t]{16.5 cm}
\caption{The superratio $R/R_{RPWIA}$ as a function of $Q^2$ in
      $^4$He. The solid curve shows RMSGA predictions using
      free-proton electromagnetic form factors. The dotted
      (dot-dashed) curve represents RMSGA calculations with in-medium
      electromagnetic form factors from the CQS (QMC) model. Data are
      from Refs.~\cite{dieterich01}(open triangle) and
      \cite{strauch03}(solid circles).}
\end{minipage}
\end{center}
\end{figure}

\end{document}